\documentclass[11pt]{article}
\usepackage{moriond,epsfig}
\usepackage[utf8]{inputenc}
\usepackage{pstricks,pst-node,pst-text,pst-3d}
\usepackage{psfrag}
\usepackage{psfig}
\usepackage{graphicx}
\usepackage{amssymb}
\usepackage{amsmath}
\usepackage{amsfonts}
\usepackage{dsfont}
\usepackage{listings}

\def\vev#1{\left\langle #1\right\rangle}

\allowdisplaybreaks

\begin{document}

\vspace*{-2cm}
\begin{flushright}
CFTP/11-011
\end{flushright}
\vspace*{4cm}
\title{LHC AND LEPTON FLAVOUR VIOLATION PHENOMENOLOGY IN SEESAW MODELS}

\author{ J.~C.~ROMÃO }

\address{Departamento de Física \& CFTP, Instituto Superior Técnico,
  Technical University of Lisbon\\
A. Rovisco Pais 1, 1049-001 Lisboa, Portugal}

\maketitle\abstracts{
We review Lepton Flavour Violation (LFV) in the supersymmetric version
of the seesaw mechanism (type I, II, III) and in Left-Right 
models. The LFV needed to explain neutrino masses and mixings
is the only source of LFV and has experimental implications both
in low-energy experiments  where we search for the radiative
decays of leptons, and at the LHC where we look at its imprint on the
LFV decays of the sparticles and on slepton mass splittings.
We discuss how this confrontation between high- and low-energy LFV
observables may provide information about the underlying mechanism of
LFV.}

\section{Introduction}

The experimental observation of non-vanishing neutrino masses and
mixings, \cite{neutrino} constitutes clear evidence for physics beyond
the Standard Model (SM).
As neutrino oscillations indisputably signal lepton flavour violation
(LFV) in the neutral sector, it is only natural to expect that charged
lepton flavour will also be violated in extensions of the SM where
$\nu$ oscillations can be naturally accommodated.
The search for manifestations of charged LFV constitutes the goal of
several experiments, \cite{LFV-low} exclusively dedicated to look for
signals of processes such as rare radiative as well as three-body
decays and lepton conversion in muonic nuclei.

In parallel to these low-energy searches, if the high-energy Large Hadron
Collider (LHC) finds signatures of supersymmetry (SUSY), it is then extremely
appealing to consider SUSY models that can also accommodate neutrino
oscillations. One of the most economical and elegant possibilities is
perhaps to embed a seesaw mechanism in this framework, the so-called
SUSY seesaw. 
 
If the seesaw is indeed the source of both neutrino masses and
leptonic mixings and accounts for low-energy LFV observables within
future sensitivity reach, we show that interesting phenomena are
expected to be observed at the LHC: in addition to measurable slepton mass
splittings, the most striking effect will be the possible appearance
of new edges in di-lepton mass distributions.

\section{Models}

\subsection{Seesaw type I,II, III \& Left-Right Model}

At GUT scale the SU(5) invariant superpotentials for type I, II and
III SUSY seesaw are \cite{papers-seesaw}
\begin{equation}
  W_{\rm RHN} = {\bf Y}_N^{\rm I}\ N^c\  \overline{5}\cdot 5_H +
  \frac{1}{2}\ M_R\  N^c N^c  \ ,
\end{equation}
\begin{align}
W_{\rm 15H}  = & \frac{1}{\sqrt{2}}{\bf Y}_{N}^{\rm II}\ {\bar 5}
\cdot 15 \cdot {\bar 5}  
   + \frac{1}{\sqrt{2}}\lambda_1 {\bar 5}_H \cdot 15 \cdot {\bar 5}_H 
+ \frac{1}{\sqrt{2}}\lambda_2 5_H \cdot \overline{15} \cdot 5_H 
+ {\bf Y}_5 10 \cdot {\bar 5} \cdot {\bar 5}_H \nonumber\\
  & + {\bf Y}_{10} 10 \cdot 10 \cdot 5_H + M_{15} 15 \cdot \overline{15} 
+ M_5 {\bar 5}_H \cdot 5_H\ ,
\end{align}
\begin{align}
W_{\rm 24 H}  = & \sqrt{2} \, {\bar 5}_M Y^5 10_M {\bar 5}_H 
          - \frac{1}{4} 10_M Y^{10} 10_M 5_H 
  +  5_H 24_M Y^{III}_N{\bar 5}_M + \frac{1}{2} 24_M M_{24}24_M \ .
\end{align}
\begin{figure}[htb]
  \centering
  \begin{tabular}{ccc}
    \psframebox[framearc=.2]{type-I} 
    &
    \psframebox[framearc=.2]{type-II}
    &
    \psframebox[framearc=.2]{type-III}
    \\[+2mm] 
    \psfrag{nL}{$\nu_L$}
    \psfrag{nR}{$N^c$}
    \psfrag{MR}{$M_R$}
    \psfrag{f}{\hskip -2mm$\vev{H_u}$}
    \includegraphics[clip,height=30mm]{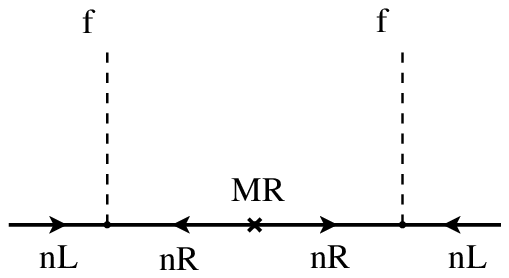}
    &
    \psfrag{nL}{$\nu_L$}
    \psfrag{nR}{$\nu_R$}
    \psfrag{MR}{$M_R$}
    \psfrag{f}{\hskip -2mm$\vev{H_u}$}
    \psfrag{D}{$T^0$}
    \psfrag{m}{$\mu$}
    \psfrag{YD}{$Y_{\nu}^{II}$}
    \includegraphics[height=30mm]{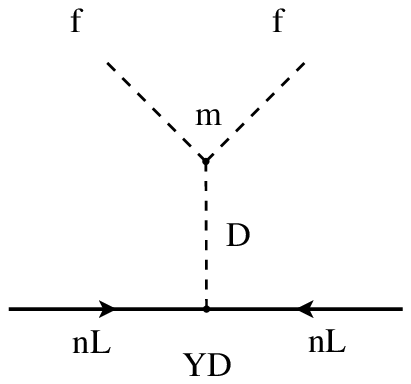}
    &
    \psfrag{nL}{$\nu_L$}
    \psfrag{nR}{$W_M$}
    \psfrag{MR}{$M_{W_M}$}
    \psfrag{f}{\hskip -2mm$\vev{H_u}$}
    \hskip -5mm
    \includegraphics[clip,height=30mm]{typeI.eps}\\[+1mm]
  \end{tabular}
  \caption{Seesaw types}
  \label{fig:1}
\end{figure}
The exchange of the singlet $N^c$ in type I, of the scalar triplet $T$ in
type II and of both the fermionic triplet $W_M$ and fermionic
singlet $B_M$ in type III lead, through the diagrams of
Fig.~\ref{fig:1} to the well known effective neutrino mass matrix
formulas, 
\begin{equation}
  m_{\rm eff}^{\rm I}= - (v Y_{\nu}) M_R^{-1} (v Y_{\nu})^T,\quad
  m_{\rm eff}^{\rm II}=  \frac{v^2 \mu Y_{\nu}^{\rm II}} {M^2_{T}},\quad
  m_{\rm eff}^{\rm III}= - (v Y_{\nu}^{\rm III}) M_{W_M}^{-1} (v
  Y_{\nu}^{\rm III})^T \ .
\end{equation}


We have also studied \cite{LR} a SUSY seesaw in which the breaking from $SU(5)$
to the SM gauge group is done in two steps, first  to a
Left-Right (LR) symmetric model, $SU(3)_c\times SU(2)_L\times
SU(2)_R\times U(1)_{B-L}$ at scale $v_R$, and then with the $B-L$ broken 
at a lower scale $v_{B-L}$. For neutrino physics, as well as for
the LFV, the relevant part of the superpotential is,  
\begin{equation}
  \label{eq:1}
 \mathcal{W}^{LR} = Y_L L \Phi L^c - f_c L^c \Delta^c L^c + \cdots \ ,
\end{equation}
where $Y_L$ and $f_c$ complex $3\times3$ matrices. After the $B-L$
breaking we have,
\begin{equation}
  \label{eq:2}
  \mathcal{L}= H_u\, \overline{\nu_L}\,Y_{\nu}^{\rm I}\, \nu_R   -
  \frac{1}{2} \nu_R^T\, C^{-1}\,  (f_c v_{BL})\, \nu_R  + \cdots \ ,
\end{equation}
leading to an effective neutrino mass matrix of the type I form,
\begin{equation}
  \label{eq:3}
m_{\rm eff}^{\rm LR}= - (v Y_{\nu}) (f_c v_{BL})^{-1} (v Y_{\nu})^T \ . 
\end{equation}
The important point here is that, as we have two complex matrices, we
can have different types of neutrino fits. We studied two limiting
situations, the so-called $Y_\nu$ fit where 
 $f_c=\mathds{1}$ ($Y_\nu$ arbitrary), and the $f$ fit
where $Y_\nu=\mathds{1}$ ($f_c$ arbitrary). These will
leave different imprints on the LFV through their RGE running.

\subsection{LFV in the Models}

Starting with universal minimal supergravity inspired (mSUGRA)
boundary conditions at $M_{\rm GUT}$, the off-diagonal entries in $Y^\nu$
will induce the LFV on the slepton mass matrices through RGE effects.
For type I, II and III we have
\begin{align}
  &\Delta m^2_{L,ij} \simeq -\frac{a_k}{8 \pi^2 } 
  \left( 3 m^2_0 +  A^2_0 \right) 
  \left(Y^{k,\dagger}_N L Y^{k}_N\right)_{ij}, \quad L =
  \ln(\frac{M_{\rm GUT}}{M_{\rm N}}) \\
  &\Delta m^2_{E,ij} \simeq 0 \hskip 20mm a_{\rm I}=1\,\, , \,\, a_{\rm II}=6
  \,\, \mathrm{and} \,\,\, a_{\rm III} =
  \frac{9}{5} \ ,
\end{align}
while for the LR model we have two situations. From $M_{\rm GUT}$ to $v_R$,
\begin{align}
  \Delta m_L^2 &\simeq 
  - \frac{1}{4 \pi^2} \left( 3 f f^\dagger + Y_L^{(k)} Y_L^{(k) 
      \: \dagger} \right) (3 m_0^2 + A_0^2) 
  \ln \left( \frac{M_{\rm GUT}}{v_R} \right) \\
  \Delta m_{E}^2 &\simeq - \frac{1}{4 \pi^2} \left( 3 f^\dagger f + Y_L^{(k) 
      \: \dagger} Y_L^{(k)} \right) (3 m_0^2 + A_0^2)
  \ln \left( \frac{M_{\rm GUT}}{v_R} \right)\ ,
\end{align}
while from $v_R$ to $v_{BL}$,
\begin{align}
  \Delta m_L^2 &\simeq  - \frac{1}{8 \pi^2} Y_\nu Y_\nu^\dagger
  \left( m_L^2|_{v_R} + A_e^2|_{v_R}\right)
  \ln \left( \frac{v_R}{v_{BL}} \right), \quad
  \Delta m_{E}^2  \simeq  0 \ .
\end{align}
Therefore, the choice of the different neutrino fits will have implications
on the lepton flavour violation observables.
The low energy LFV processes are described by an effective Lagrangian, 
\begin{equation}
  \mathcal{L}_{eff} = e \frac{m_{l_{i}}}{2}\ 
\bar{l}_i \sigma_{\mu \nu} F^{\mu \nu} 
  (A_L^{ij} P_L + A_R^{ij} P_R) l_j + h.c. 
\end{equation}
For seesaw models,
\begin{equation}
  \label{eq:4}
  A_L^{ij} \sim \frac{(\Delta m_L^2)_{ij}}{m_{\rm SUSY}^4}, \quad A_R^{ij} 
  \sim \frac{(\Delta m_{E}^2)_{ij}}{m_{\rm SUSY}^4}\ . 
\end{equation}
This implies that for type I, II and II we have only $A_L\not=0$, while for
the LR model we can have both, $A_L$ and $A_R$. This implies that if
MEG \cite{LFV-low} finds evidence for the decay $\mu^+ \to e^+ \gamma$, then
we can distinguish among the models
by looking at the positron polarization asymmetry,
\begin{equation}
  \label{eq:5}
    \mathcal{A}(\mu^+ \to e^+ \gamma) =
    \frac{|A_L|^2-|A_R|^2}{|A_L|^2+|A_R|^2}\ \left\{
      \begin{array}{cl}
        = 1 &\text{type-I-II-III}\\
        \not= 1 &\text{LR}
      \end{array}
    \right.  \quad .
\end{equation}

\section{Results}

For all the models we have
studied~\cite{papers-seesaw}${}^{\!,\,}$\cite{LR}${}^{\!,\,}$\cite{Abada:2010kj}
the different low- and high-energy LFV observables.  The numerical
analysis was done using the public code SPheno, \cite{SPheno} that
includes the 2-loop RGEs calculated with the public code
SARAH. \cite{Staub:2008uz} 

\subsection{Low-Energy Observables}

\begin{figure}[htb]
  \centering
  \begin{tabular}{cc}
    \includegraphics[width=0.45\textwidth]{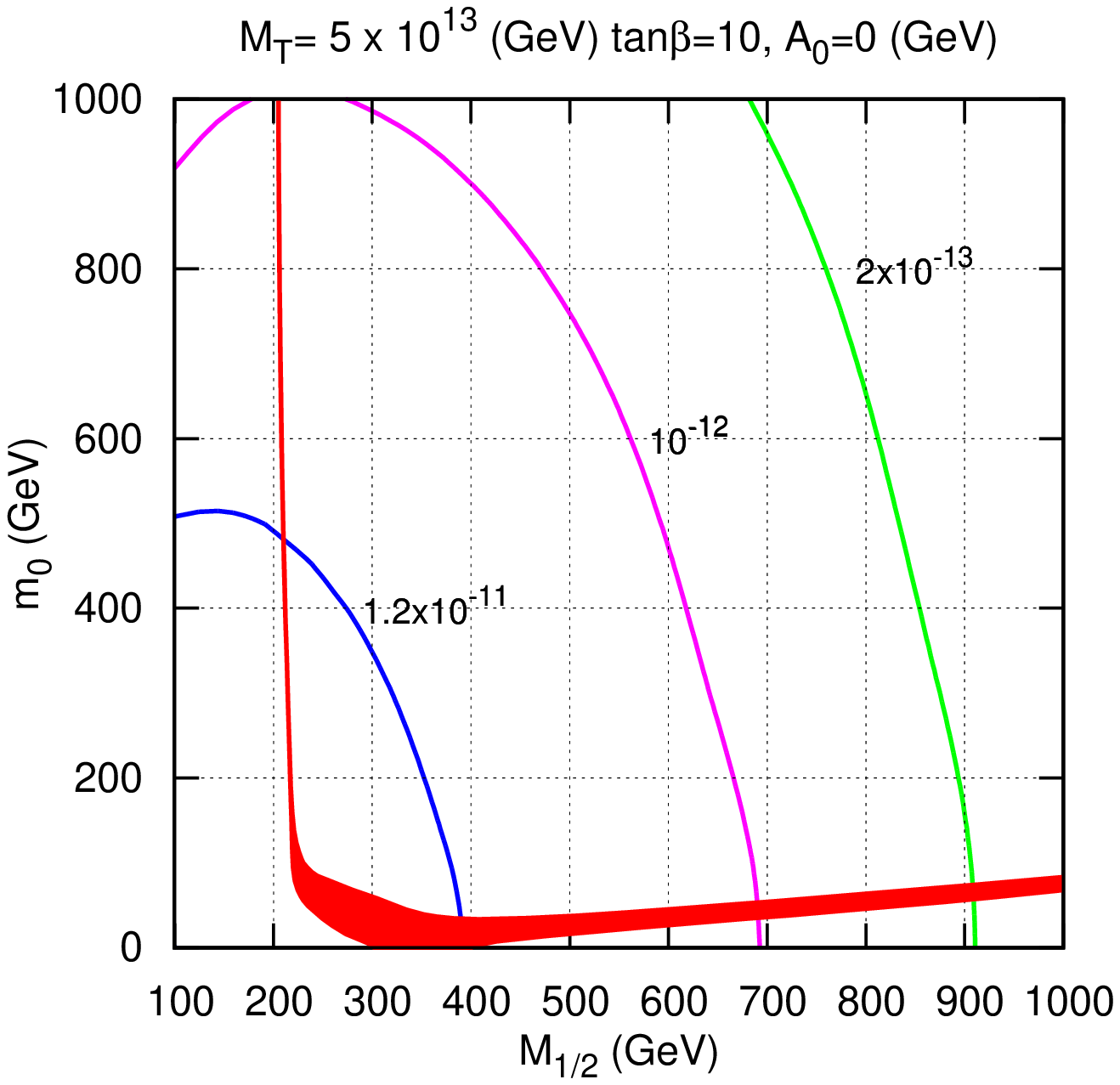}&
    \includegraphics[width=0.45\textwidth]{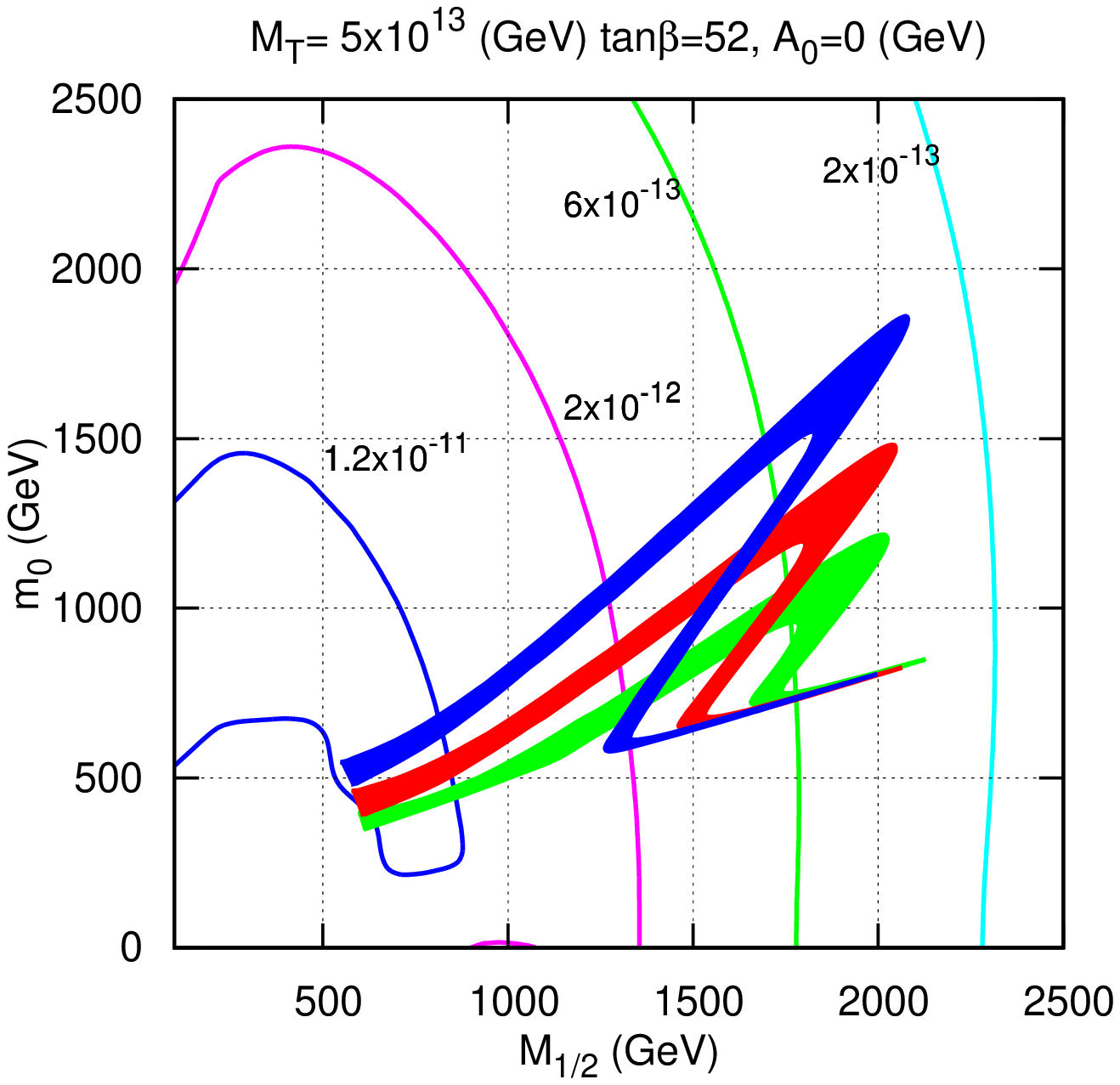}
  \end{tabular}
\vspace{-3mm}
  \caption{Dark matter allowed regions and BR($\mu\to e \gamma$)
    contours for type II,}
  \label{fig:4}
\end{figure}

The present bounds on low-energy LFV observables and dark matter
abundance already constrain the parameter space of the models. As an
example we give in Fig.~\ref{fig:4} the type II case. On the left
panel we show the allowed regions for dark matter abundance (within
3$\sigma$ of the WMAP~\cite{WMAP} observation). A scan was performed
in the $M_{1/2}-m_0$ plane, the other cMSSM parameters being taken as
$A_0=0$, $\tan\beta=10$, $\mu>0$. The seesaw scale was
fixed at $M_T= 5\times 10^{13}$ GeV. Superimposed are the contours for
BR($\mu\to e \gamma$). We see that for these input parameters only a
small part of the parameter space remains viable after imposing the LFV and
dark matter constrains. Once MEG gets to the sensitivity of $10^{-13}$,
most of the parameter space will be excluded if no signal is found. On
the right panel of Fig.~\ref{fig:4} we show a similar plot, now in the
so-called Higgs funnel region obtained for $\tan\beta=52$, the other
parameters as before. The variation with the top mass is shown:
$m_{top}=169.1$ GeV (blue), $171.2$ GeV (red), $173.3$ GeV (green).

\begin{figure}[htb]
  \centering
  \begin{center}
  \includegraphics[width=0.43\textwidth]{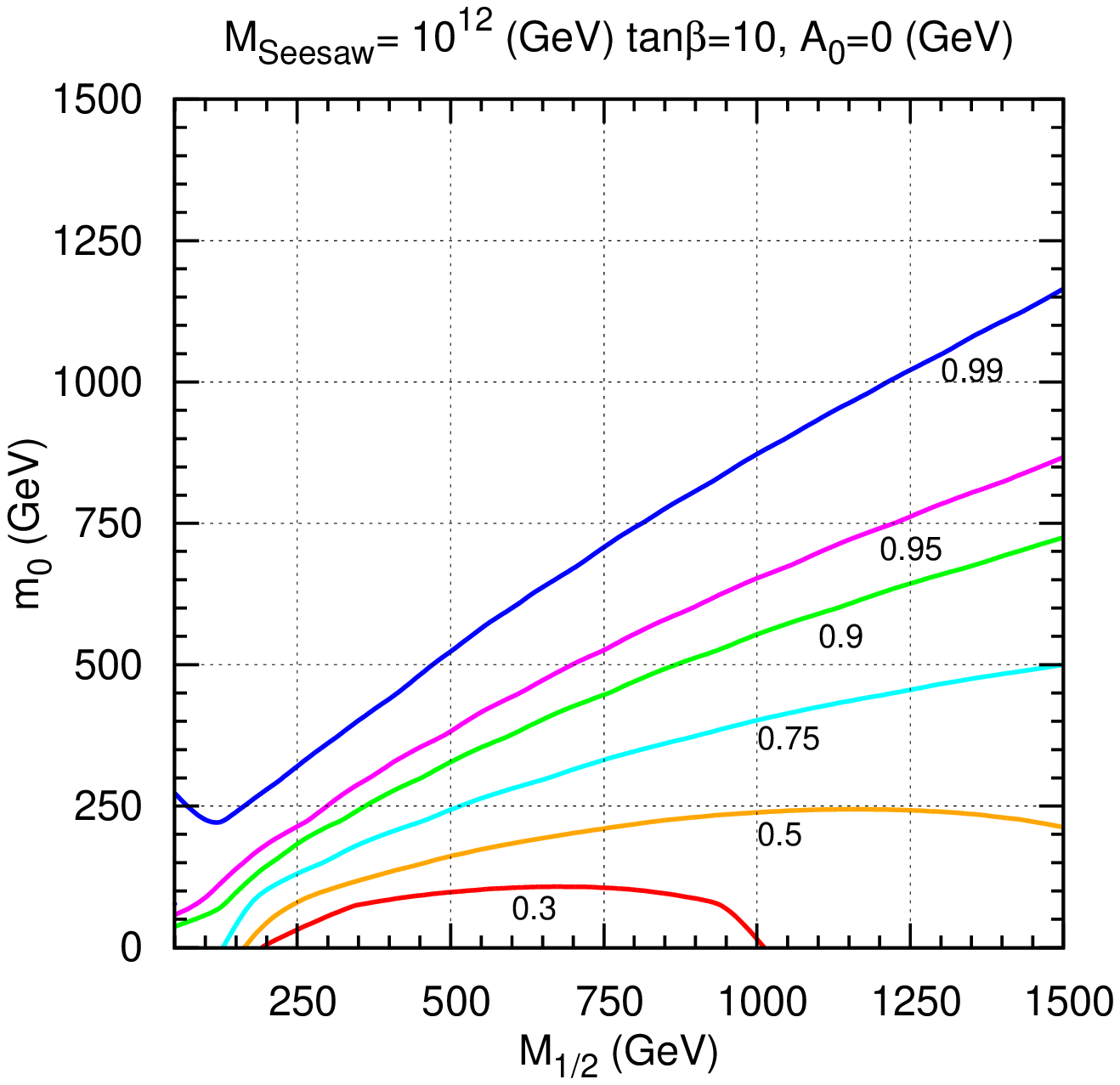}
  \includegraphics[bb=100 400 470 700, width=0.43\textwidth]{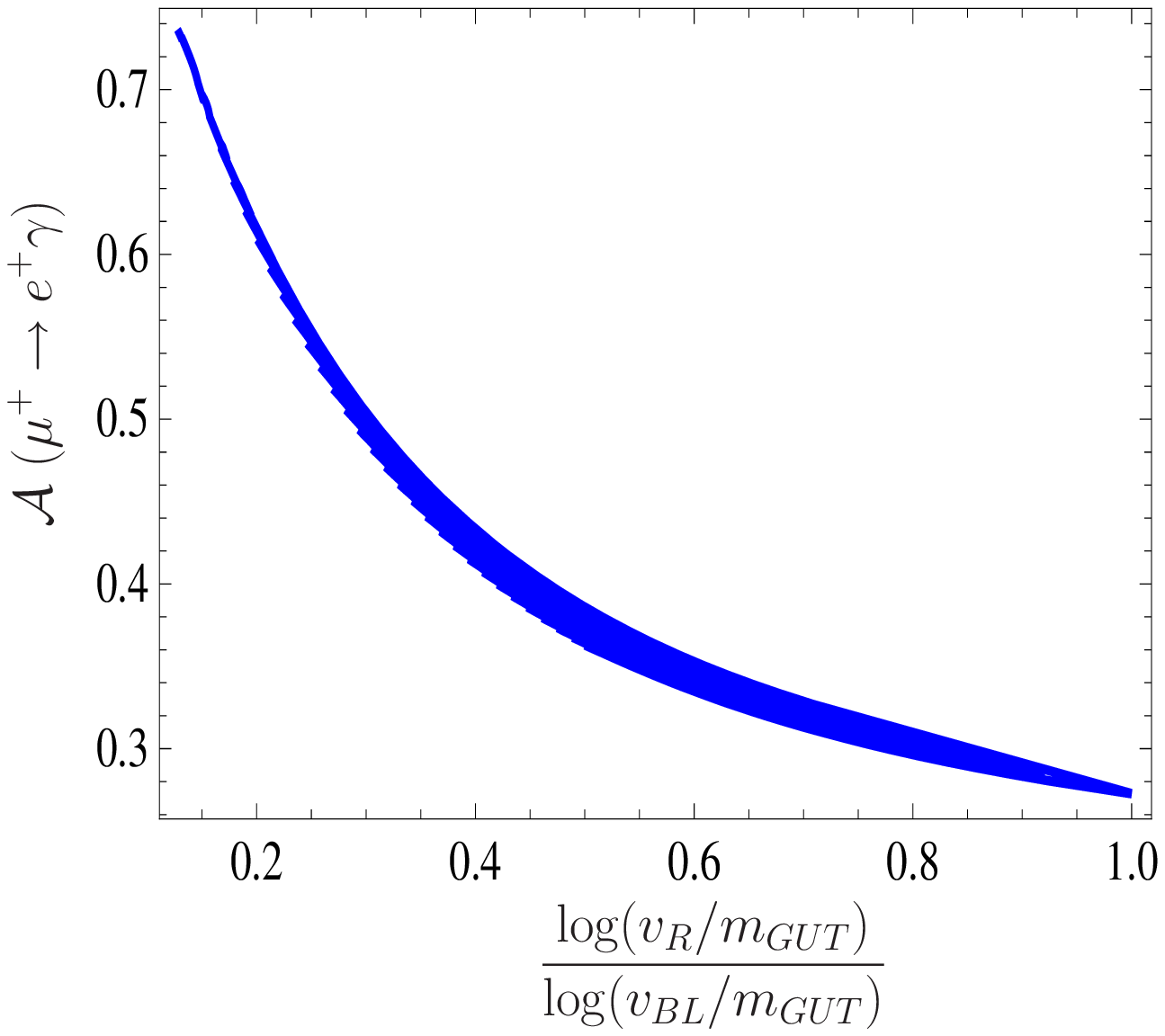}
  \end{center}
\vspace{-2mm}
  \caption{Positron asymmetry in the Left-Right model.}
  \label{fig:5}
\end{figure}

As another example we consider the $e^+$ asymmetry defined in
Eq.~(\ref{eq:5}) in the LR model~\cite{LR}. On the
left panel of Fig.~\ref{fig:5} we show the contours for $\mathcal{A}$
in the $M_{1/2}-m_0$ plane. The cMSSM parameters were taken as those of
the SPS3 point, $m_0=90$ GeV, $M_{1/2}=400$ GeV, $A_0=0$ GeV,
$\tan\beta=10$ and $\mu>0$. We take
$M_{\rm Seesaw}=10^{12}$ GeV, while the LR breaking scales were
$v_{BL}=10^{15}$ GeV, $v_R \in [10^{14},10^{15}]$ GeV and 
 $Y_\nu$ fit was chosen. On the right panel we show, for the same
parameters, the correlation between the asymmetry and the breaking
scales. If MEG measures $\mathcal{A} < 1$, we can have an handle on
the scales $v_R,v_{BL}$ and test the LR model.

\subsection{LHC Observables}

\begin{figure}[htb]
  \centering
  \begin{tabular}{ll}
    \hskip -65mm
  \includegraphics[width=0.52\textwidth]{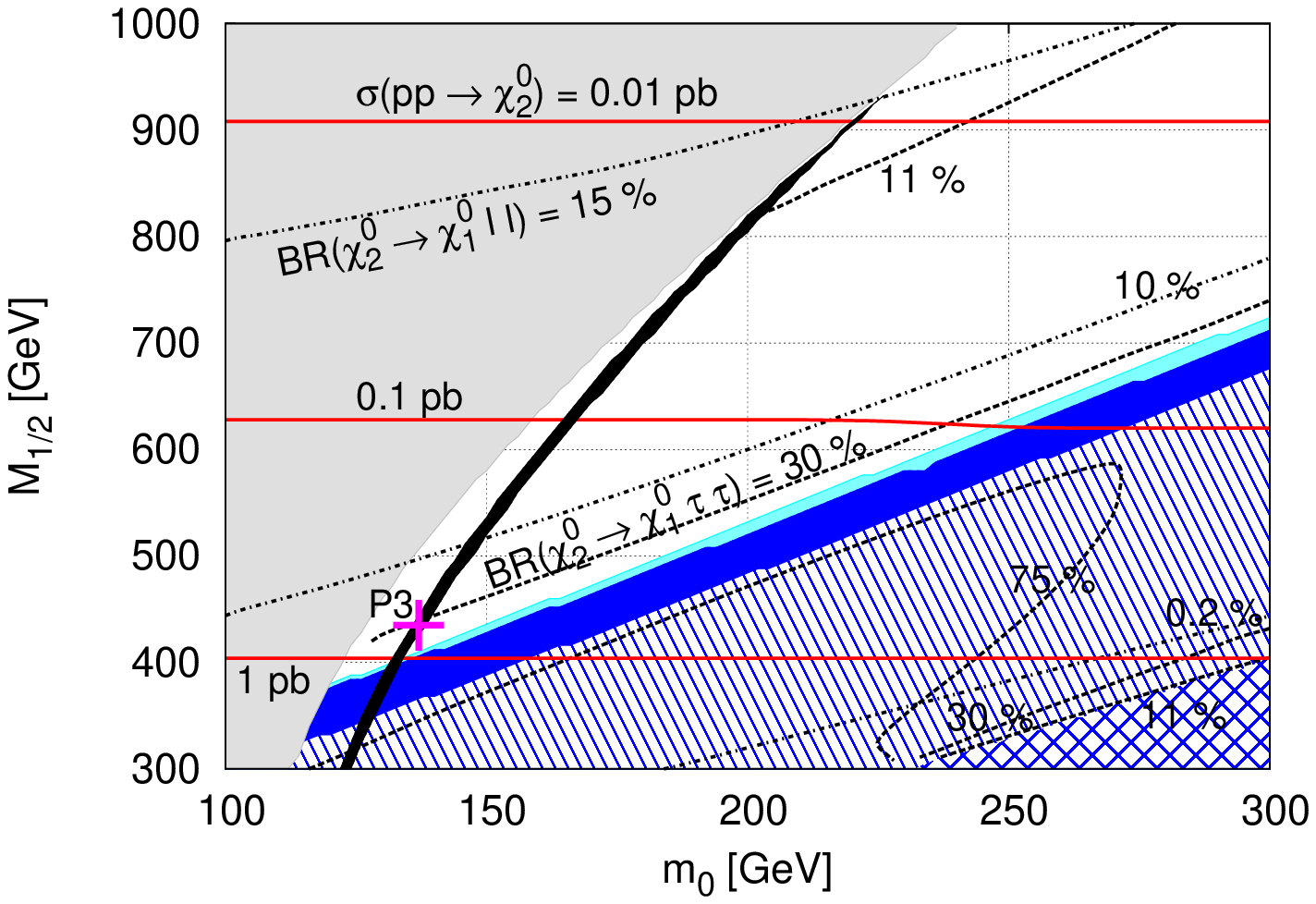}&    
  \rput(3,3.2){\scalebox{0.85}{\begin{tabular}{|c|c|c|c|c|}
    \hline
    Point & $m_0$ & $M_{1/2}$ & $A_0$ & $\tan \beta$ \\
    &  (GeV)&  (GeV)& (TeV)& \\
    \hline
    P1& 110 & 528& 0& 10\\
    \hline 
    P2& 110& 471& 1& 10\\
    \hline 
    P3& 137& 435& -1& 10\\
    \hline 
    P4& 490& 1161& 0& 40\\
    \hline 
    P5-HM1& 180&850 &0 &10 \\
    \hline 
    P6-SU1& 70& 350& 0& 10\\
    \hline
  \end{tabular}}}
  \end{tabular}
\vspace{-5mm}
  \caption{Standard window (see text) and benchmark points
  used in the analysis.}
  \label{fig:6}
\end{figure}

\begin{figure}[htb]
  \centering
    \includegraphics[width=0.9\linewidth]{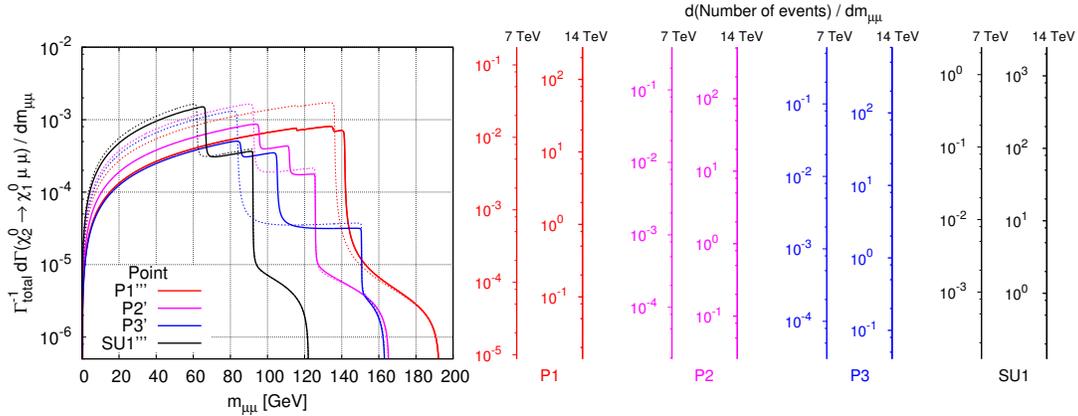}
\vspace{-3mm}
  \caption{Di-muon invariant mass distribution for the SUSY seesaw for the
    benchmark points defined in Fig.~\ref{fig:6}.}
  \label{fig:8}
\end{figure}

\begin{figure}[htb]
   \centering
  \begin{tabular}{cc}
    \includegraphics[width=0.45\linewidth,clip]{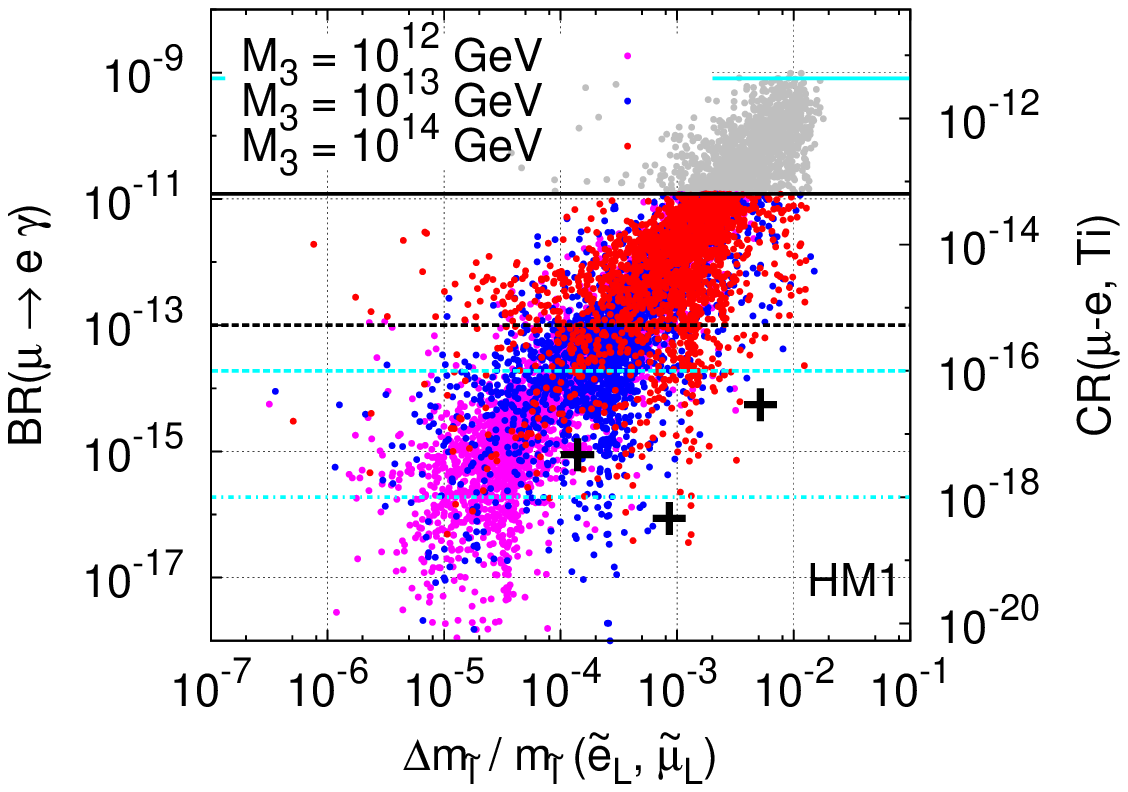}&
    \hskip -2mm
    \includegraphics[width=0.45\linewidth,clip]{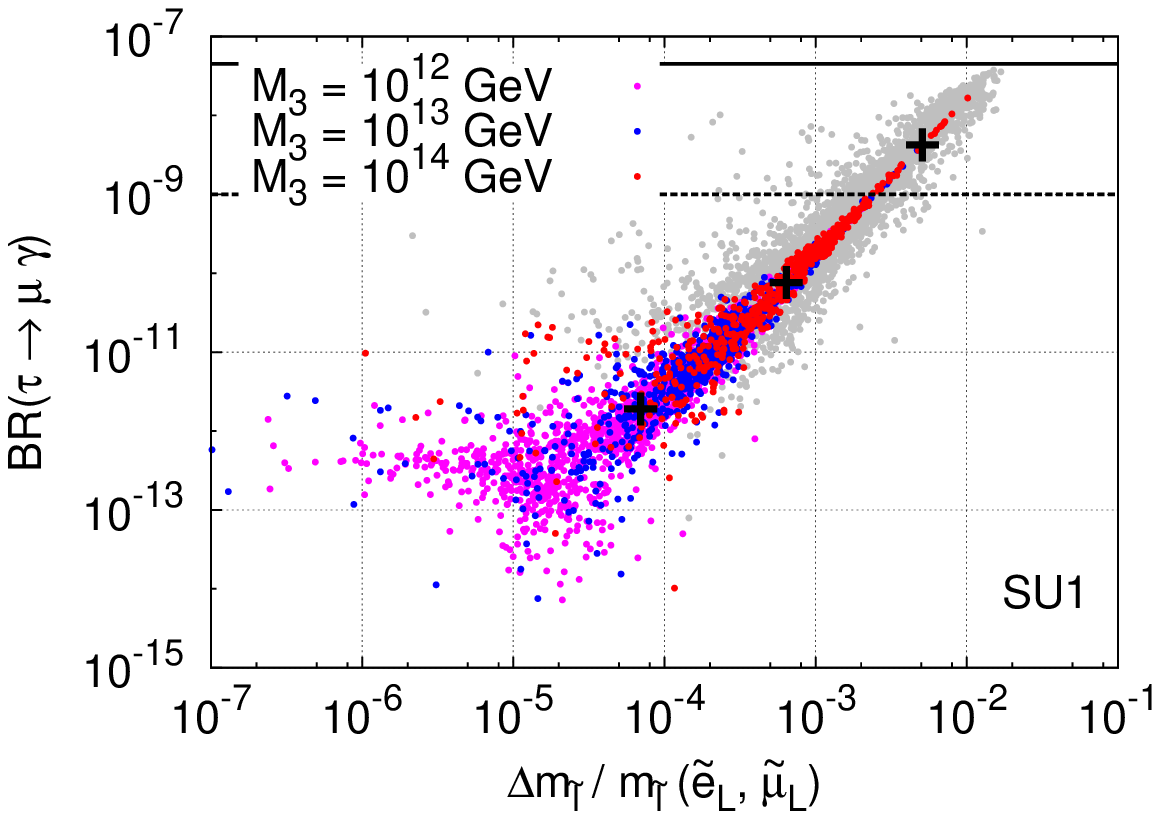}
  \end{tabular} 
\vspace{-6mm}
   \caption{Correlation between low-energy and LHC observables for the
   benchmark points HM1 and SU1.}
   \label{fig:9}
 \end{figure}

At LHC we look at di-lepton invariant mass distributions from
  $\chi_2^0 \rightarrow \tilde{\ell}^i_{L,R} \ell_ \rightarrow
  \chi_1^0 \ell \ell$ decays,
that can be measured with a precision of $0.1\%$, \cite{MS}
for on-shell sleptons and isolated leptons with large $p_T > 10$ GeV. 
From this we can infer the slepton mass splittings, 
\begin{equation}
  \label{eq:7}
  \frac{\Delta m_{\tilde \ell}}{m_{\tilde \ell}} (\tilde \ell_i, \tilde
  \ell_j) \, = \, 
\frac{|m_{\tilde \ell_i}-m_{\tilde \ell_j}|}{<m_{\tilde
    \ell_{i,j}}>}\quad @\text{LHC}:
 \begin{array}{l}
   \Delta m/m_{\tilde \ell} (\tilde e_L,\tilde\mu_L)\sim
   \mathcal{O}(0.1\%)\\
   \Delta m/m_{\tilde \ell} (\tilde \mu_L,\tilde\tau_L) \sim
   \mathcal{O}(1\%)\ .
 \end{array}
\end{equation}
We start our analysis by identifying what we call a
\textit{standard window}. This is defined by the requirement of having
on-shell sleptons decaying with isolated leptons with large $p_T > 10$
GeV. We also require large $\chi_2^0$ production, a sizable
BR($\chi_2^0\rightarrow \chi_1^0\ell \ell$) and, if possible, the
correct abundance of dark matter, $\Omega h^2$. This is shown on the
left panel of Fig.~\ref{fig:6},
where the white region fulfills all the requirements (the correct dark
matter abundance corresponds to the black line inside the region). To
carry out our analysis we chose the cMSSM study points shown in the
right panel of Fig.~\ref{fig:6} and then varied the seesaw
parameters. In the cMSSM we get double-triangular distributions
corresponding to intermediate $\tilde\mu_L$ and $\tilde\mu_R$ in
$\chi_2^0\!\rightarrow\!  \chi_1^0\mu\mu$, with superimposed
$\tilde\ell_{L,R}$ edges for $m_{\mu\mu}$ and $m_{ee}$ because of
``degenerate'' $\tilde\mu,\tilde e$.
In Fig.~\ref{fig:8} we show the di-muon invariant distribution, and
number of expected events, for the case of SUSY type I seesaw, for the
following choice of seesaw parameters: $M_N$=$\{10^{10}, 5 \times
10^{10}, 5\times 10^{13}\}$ GeV (P2$'$, P3$'$) and $M_N$=$\{10^{10}, 5 \times
10^{12}, 10^{15}\}$ GeV (P1$'''$, SU1$'''$), always with
$\theta_{13}=0.1^{\circ}$.
We get displaced $m_{\mu\mu}$ and $m_{ee}$ edges ($\ell_L$) which give
sizable mass splittings $\frac{\Delta m_{\tilde
    \ell}}{m_{\tilde\ell}}(\tilde e_L,\tilde\mu_L)$. We also find the
appearance of a new edge in $m_{\mu\mu}$ corresponding to an
intermediate $\tilde\tau_2$. These mass splittings are correlated with
the low-energy observables as we show in Fig.~\ref{fig:9}. On the left
panel we show the correlation for BR($\mu\to e \gamma$) for the CMS
benchmark point HM1 ($m_0=180$ GeV, $M_{1/2}=800$ GeV, $A_0=0$ GeV,
$\tan\beta=10$ and $\mu>0$)
while on the right panel we show the correlation for  BR($\tau\to \mu
\gamma$) for the ATLAS benchmark point SU1 ($m_0=70$ GeV,
$M_{1/2}=350$ GeV, $A_0=0$ GeV, $\tan\beta=10$ and $\mu>0$). In these
plots we performed a scan over the SUSY seesaw parameters, with
 $M_{N_{3}}=10^{12,13,14}$ GeV,  $\theta_{13}=0.1^{\circ}$.

We conclude that if SUSY is discovered with a spectrum similar to HM1
or SU1 and a type-I seesaw is at work, then the LFV observables will
be within experimental reach at LHC, while BR($\mu\rightarrow
e\gamma$) and BR($\tau\rightarrow e\gamma$) will be within the reach of
MEG and SuperB, respectively.

\section{Conclusions}

In SUSY seesaw models the neutrino Yukawa couplings, $Y_\nu$, acts as the only
source of LFV, implying a correlation between low- and high-energy
LFV observables. We have performed a study of these correlations in
the so-called SUSY seesaws type I, II and III, as well as in a seesaw
model that  is Left-Right symmetric below the GUT scale.
  
If SUSY seesaw is to account for neutrino masses and mixings then we
will have slepton mass splittings within
LHC sensitivity, with the possible observation of new edges in the di-lepton
invariant mass distributions. In most cases a clear correlation can be
established between low- and high-energy LFV observables
(e.g. BR vs $\Delta m_{\tilde \ell}$) due to their unique
source.
 
The experimental data that will be available soon, both from the high-
and low-energy experiments, will either substantiate the seesaw
hypothesis, or disfavour the SUSY seesaw as the (only) source of flavour
violation.

\section*{Acknowledgments}

 This work has been done partly under 
the EU Network grant UNILHC PITN-GA-2009-237920 and from {\it
 Funda\c{c}\~ao para a Ci\^encia e a Tecnologia} grants CFTP-FCT UNIT
 777,  PTDC/FIS/102120/2008, CERN/FP/109305/2009.

\end{document}